\documentclass{PoS}

\title{Electromagnetic structure of the low-lying baryons in covariant chiral perturbation theory}

\ShortTitle{Electromagnetic structure of the low-lying baryons in covariant chiral perturbation theory}

\author{\speaker{Jorge Martin Camalich}\\
        IFIC Valencia\\
        E-mail: \email{camalich@ific.uv.es}}
\author{Luis Alvarez-Ruso\\
Departamento de F\'isica, Universidade de Coimbra
      \\
        E-mail: \email{luis.alvarez@ific.uv.es}}
\author{Lisheng Geng\\
     IFIC Valencia\\
        E-mail: \email{lsgeng@ific.uv.es}}

\author{Manuel Jose Vicente Vacas\\
     IFIC Valencia\\
        E-mail: \email{vicente@ific.uv.es}}

\abstract{We report a calculation of the low-lying baryon magnetic moments using covariant chiral perturbation theory within the
 extended-on-mass-shell renormalization scheme including intermediate octet and decuplet contributions. For the case of
 the baryon octet, we succeed to improve the Coleman-Glashow description by including the leading SU(3)$_F$-breaking effects 
 coming from the lowest-order loops. We compare with previous attempts at the same order 
using heavy-baryon and covariant infrared chiral perturbation theory, and discuss the source of the differences. For the 
case of the decuplet-baryons we fix the only unknown LEC with the well measured magnetic dipole moment of the $\Omega^-$
and predict the corresponding ones of the $\Delta(1232)$ isospin multiplet. In particular we obtain
$\mu_{\Delta^{++}}=6.0(6)\;\mu_N$ and $\mu_{\Delta^{+}}=2.84(34)\;\mu_N$ that compare well with the current experimental
information.}

\FullConference{6th International Workshop on Chiral Dynamics, CD09\\
		July 6-10, 2009\\
		Bern, Switzerland}

\begin{document}

\section{Introduction}

The magnetic moments of the baryons are of utmost importance since they contain information on their internal structure as read out by electromagnetic probes or photons. A starting point in the understanding of the magnetic moments of 
the octet is the SU(3)$_F$-symmetric model of Coleman and Glashow~\cite{Coleman:1961jn} that describes the eight magnetic moments and the
$\Lambda\Sigma_0$ transition moment in terms of two parameters. The success of this model relies on the almost 
preserved global SU(3)$_F$-symmetry of QCD with $u$, $d$ and $s$ flavors. The description of the symmetry-breaking corrections of the baryon magnetic moments
can be addressed in a systematic and model-independent fashion by means of chiral perturbation theory ($\chi$PT)
~\cite{Gasser:1984gg,Gasser:1987rb}. Different calculations have been performed in the last decades either in the heavy-baryon (HB)
~\cite{Caldi:1974ta,Jenkins:1990jv,Jenkins:1992pi,Meissner:1997hn,Durand:1997ya,Puglia:1999th}, covariant infrared (IR)
~\cite{Becher:1999he,Kubis:2000aa} or covariant extended-on-mass-shell (EOMS)~\cite{Fuchs:2003qc,Geng:2008mf,Geng:2009hh} versions
of baryon $\chi$PT. Only in the EOMS approach a good convergence of the chiral series and an improvement over the
Coleman-Glashow description at next-to-leading-order is obtained. The comparison with the other $\chi$PT schemes highlights the importance of analyticity and
full covariance~\cite{Geng:2008mf} and the relevance of implementing the spin-3/2 resonances
within a \textit{consistent} framework~\cite{Pascalutsa:2000kd,Pascalutsa:2006up,Geng:2009hh}. 

The aforementioned covariant approach that includes both octet and decuplet contributions have been also
applied to the description of the electromagnetic structure of the decuplet resonances~\cite{Geng:2009ys} updating the HB
calculation of Ref.~\cite{Butler:1993ej}. In particular, the
magnetic dipole moments of the $\Delta^+$ and $\Delta^{++}$ are predicted using the well-measured one of the $\Omega^-$ to
fix the only LEC appearing up to this order. The relevance of these results lies on the ongoing efforts from the
experimental side to measure the magnetic moments of these two resonances (see for instance Refs.~\cite{Machavariani:1999fr,Kotulla:2002cg,Kotulla:2008zz}). On the theoretical side,
calculations from many different approaches have arisen in the last decades (see for instance~\cite{Geng:2009ys} for a complete list
of references). In particular, the lattice QCD have recently made
remarkable progress to tackle the study of these observables from first principles~\cite{Alexandrou:2008bn,Aubin:2008qp,
Boinepalli:2009sq}. 

Finally it is worth mentioning that this covariant formalism has been applied to predict, up to $\mathcal{O}(p^4)$, 
the vector hyperon decay charge $f_1(0)$~\cite{Geng:2009ik}, which is essential to extract the Cabibbo-Kobayashi-Maskawa
matrix element $V_{us}$ from the hyperon decay data.

\section{Results}

\begin{figure}[t]
\includegraphics[width=\columnwidth]{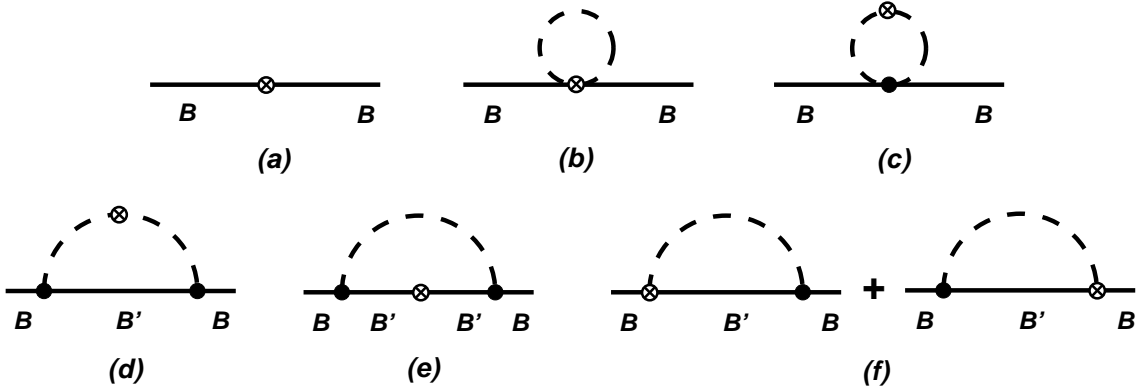}
\caption{Feynman diagrams contributing to the electromagnetic form factors of
the octet and decuplet baryons. The external and internal lines represent either
octet or decuplet baryons, the dots lowest-order meson-baryon-baryon vertices
and the crossed dot the coupling of the external electromagnetic current.   \label{Fig:diagrams}}
\end{figure}

The Feynman diagrams contributing to the electromagnetic form factors of the
octet and decuplet baryons are displayed in Fig.~(\ref{Fig:diagrams}). All the
vertices are obtained from the chiral Lagrangians shown in Refs.~\cite{Geng:2008mf,Geng:2009hh,Geng:2009ys}. In particular,
for the magnetic moments,
the vertices describing the tree-level diagram \textbf{\textit{(a)}} introduce two unknown parameters, the LECs $b_6^D$ and
$b_6^F$, for the octet and one unknown parameter, the LEC $g_d$, for the decuplet. These
describe the leading order (LO) SU(3)$_F$-symmetric contribution to the baryon-octet and -decuplet magnetic moments respectively. If the
two LECs appearing for the baryon-octet are then fitted to the corresponding experimental values, one recovers the description
of the Coleman-Glashow model. For the case of the decuplet-resonances, the single unknown LEC can be fixed using the well
measured magnetic moment of the $\Omega^-$, $\mu_{\Omega^-}=-2.02\pm0.05$. At next-to-leading order (NLO) the only
contributions are given by the loop-diagrams
\textbf{\textit{(b)}}-\textbf{\textit{(f)}} that
depend on relatively well known couplings and masses (see Refs.~\cite{Geng:2008mf,Geng:2009hh,Geng:2009ys} for details).
Therefore, the SU(3)$_F$-breaking of the magnetic moments up to NLO comes as a prediction in $\chi$PT.

Nonetheless, the $\overline{MS}$-regularized loops with heavy fields as baryons produce power-counting problems as first noticed
in~\cite{Gasser:1987rb}. Different techniques to recover the counting have been developed, the most important of which are
the HB~\cite{Jenkins:1990jv}, the covariant IR~\cite{Becher:1999he} and covariant EOMS~\cite{Fuchs:2003qc} renormalization 
prescriptions. All of them produce loop-amplitudes with the same non-analytic terms at a given order. The covariant approaches
besides include relativistic corrections, although only the EOMS implement them in a way consistent with analyticity (see for
instance~\cite{Holstein:2005db}).

\begin{table*}
\centering
\caption{Baryon octet magnetic moments in chiral perturbation theory up to $\mathcal{O}(p^3)$. We compare the 
SU(3)-symmetric description with the different $\mathcal{O}(p^3)$ $\chi$PT calculations discussed in the text. 
We display the HB and the EOMS-covariant results with (O+D) and without (O) the inclusion of dynamical decuplets 
and the IR-covariant with only dynamical octet-baryons. We also include the experimental values from Ref.~\cite{Amsler:2008zzb}. 
\label{Table:Results}}
\begin{tabular}{c|c|cc|c|cc|c|}
\cline{2-8}
& &\multicolumn{2}{|c|}{Heavy Baryon $\mathcal{O}(p^3)$}&Cov. IR $\mathcal{O}(p^3)$&\multicolumn{2}{|c|}{Cov. EOMS $\mathcal{O}(p^3)$}& \\
\cline{3-7}
&  \raisebox{1ex}[0pt]{$\mathcal{O}(p^2)$}& O & O+D& O &O &O+D & \raisebox{1ex}[0pt]{Expt.}\\ 
\hline\hline
\multicolumn{1}{|c|}{\textit{p}} & 2.56 &3.01& 3.47 &2.25& 2.60 & 2.61& 2.793(0)\\
\multicolumn{1}{|c|}{\textit{n}} & -1.60 &-2.62&-2.84& -2.67&-2.16 &-2.23 &-1.913(0)\\
\multicolumn{1}{|c|}{$\Lambda$}& -0.80 &-0.42&-0.17&-0.61&-0.64&-0.60&-0.613(4) \\
\multicolumn{1}{|c|}{$\Sigma^-$}& -0.97 &-1.35&-1.42&-1.15& -1.12 &-1.17&-1.160(25)\\
\multicolumn{1}{|c|}{$\Sigma^+$}& 2.56 &2.18&1.77&2.38& 2.41&2.37&2.458(10) \\
\multicolumn{1}{|c|}{$\Sigma^0$}& 0.80 &0.42&0.17&0.61& 0.64 & 0.60 & ... \\
\multicolumn{1}{|c|}{$\Xi^-$}& -1.60 &-0.70&-0.41&-1.12& -0.93 & -0.92 & -0.651(3) \\
\multicolumn{1}{|c|}{$\Xi^0$}&-0.97 &-0.52&-0.56&-1.28& -1.23 & -1.22 & -1.250(14) \\
\multicolumn{1}{|c|}{$\Lambda\Sigma^0$} & 1.38 &1.68&1.86& 1.88 &1.58&1.65& $\pm$1.61(8)\\
\hline
\multicolumn{1}{|c|}{$b_6^D$}& 2.40&4.71&5.88 &4.70& 3.92 & 4.30 & \\
\multicolumn{1}{|c|}{$b_6^F$}& 0.77 &2.48&2.49&0.43& 1.28 & 1.03 & ---\\
\multicolumn{1}{|c|}{$\chi^2$}&0.46&1.01&2.58&1.30& 0.18 & 0.22& \\
\hline
\end{tabular}
\end{table*}

The question is then whether the SU(3)$_F$-breaking corrections to the baryon-octet magnetic moments can be successfully
addressed from a first principle approach by means of $\chi$PT; namely whether or not the chiral loops improve
the Coleman-Glashow description. In Table \ref{Table:Results} we show the numerical results for the baryon magnetic moments obtained by minimizing 
$\chi^2\equiv\sum(\mu_{th}-\mu_{expt})^2$ as a function of the abovementioned two unknown LECs. We have not included the 
$\Lambda\Sigma^0$ transition moment in the fit and, therefore, it is a prediction.  We compare the SU(3)-symmetric 
description with different $\mathcal{O}(p^3)$ $\chi$PT approaches. Namely, we display the HB 
and the EOMS-covariant results with (O+D) and without (O) the inclusion of dynamical-decuplets states and the IR-covariant 
with only dynamical octet-baryons. The experimental values are obtained from Ref.~\cite{Amsler:2008zzb}. 

For the HB approach, one sees how the corrections of the dynamical baryon -octet and -decuplet go in the same direction 
and are of equivalent size. Consequently, the description obtained with only the baryon-octet, that already overestimated 
the SU(3)-breaking corrections, gets much worsened when the decuplet is included. From the results obtained in HB, one is 
unavoidably led to wonder about the contributions of higher-mass resonances. In the covariant approach we have different
results attending to the renormalization scheme or to the inclusion of the decuplet. Restricting ourselves first to the 
comparison between the EOMS and IR schemes with only dynamical baryon-octet 
contributions, we observe large numerical discrepancies for differences that are formally of higher-order. As it was 
explained in Ref.~\cite{Geng:2008mf}, we interpret the unrealistic IR behaviour as a manifestation of the alteration 
that this approach produces on the analytical properties of the loop-functions. Our results indicate that this known 
problem of the IR amplitudes may already be relevant for the scales around the masses of $K$ and $\eta$ mesons. 

\begin{table*}
\centering
\caption{Values in nuclear magnetons ($\mu_N$) of the magnetic dipole moments of the $\Delta(1232)$ isospin multiplet
in SU(3)$_F$ relativistic chiral perturbation theory up to $\mathcal{O}(p^3)$ including both octet and decuplet loop
contributions (O+D). We compare our results with the 
SU(3)-symmetric prediction and with the experimental values  presented by the Particle Data Group~\cite{Amsler:2008zzb}.
\label{Table:ResMDM}}
\begin{tabular}{c|cccc|}
\cline{2-5}
&$\Delta^{++}$&$\Delta^+$&$\Delta^0$&$\Delta^-$\\
\hline
\multicolumn{1}{|c|}{SU(3)-symm.}&4.04&2.02&0&-2.02\\
\hline
\multicolumn{1}{|c|}{Covariant(O+D)}&6.0(6)&2.84(34)&-0.36(11)&-3.56(11)\\
\hline
\multicolumn{1}{|c|}{Expt.}&5.6$\pm$1.9&$2.7^{+1.0}_{-1.3}\pm1.5\pm3$&---&---\\
\hline
\end{tabular}
\end{table*}

The results in EOMS show an unprecedented NLO improvement over the tree-level description within dimensionally regularized 
$\chi$PT, where octet and decuplet contributions have been included. We can study the convergence properties of the chiral 
series factorizing the tree-level at $\mathcal{O}(p^2)$ from the whole result up to $\mathcal{O}(p^3)$. We also separate 
the loop fraction into the octet (second number) and the decuplet (third number) parts in the parenthesis
\begin{eqnarray*}
&&\mu_p=3.46(1-0.28+0.035)\hspace{0.15cm},\hspace{0.15cm}\mu_n=-2.86(1-0.16-0.06)\hspace{0.15cm},\hspace{0.15cm}\mu_\Lambda=-1.43(1-0.46-0.12),\\
&&\mu_{\Sigma^-}=-0.60(1+0.25+0.70)\hspace{0.15cm},\hspace{0.15cm}\mu_{\Sigma^+}=3.46(1-0.34+0.025)\hspace{0.15cm},\hspace{0.15cm}\mu_{\Sigma^0}=1.41(1-0.47-0.11),\\
&&\mu_{\Xi^-}=-0.60(1-0.07+0.61)\hspace{0.15cm},\hspace{0.15cm}\mu_{\Xi^0}=-2.86(1-0.48-0.09)\hspace{0.15cm},\hspace{0.15cm}\mu_{\Lambda\Sigma^0}=2.48(1-0.28-0.06).\label{Eq:convergence}
\end{eqnarray*}

Except for the $\Sigma^-$, the relative contributions of the octet and the decuplet and the overall $\mathcal{O}(p^3)$ 
terms, are consistent with a maximal correction of about $m_\eta/\Lambda_{\chi SB}$, and the decuplet corrections 
are, in general, smaller than the octet ones. Moreover, in Ref.~\cite{Geng:2009hh} the large decuplet mass limit was 
investigated and it was found that the decuplet decouples for an average $M_D$ slightly above the physical average.  

For the decuplet resonances only the magnetic moment of one of the members of the decuplet, namely the $\Omega^-$, 
has been accurately measured. In Table~\ref{Table:ResMDM} we show the predicitons in nuclear magnetons ($\mu_N$) for the 
$\Delta(1232)$ isospin multiplet magnetic moments in SU(3)$_F$ relativistic chiral perturbation theory up to 
$\mathcal{O}(p^3)$. We compare our results with the SU(3)-symmetric description and with the experimental values 
presented by the Particle Data Group~\cite{Amsler:2008zzb}. The error bars are an estimation of higher-order contributions
obtained looking at the ratio between NLO and LO contributions and assuming a good convergence of the chiral 
series (we take $30\%$ of the NLO over LO ratio)~\cite{Geng:2009ys}. 

Our results are consistent with the central value of the experimental numbers for $\mu_{\Delta^{++}}$ and 
$\mu_{\Delta^+}$. Furthermore, for the former we do agree very satisfactorily with the latest experiment, 
$\mu_{\Delta^{++}}=6.14\pm0.51$~\cite{LopezCastro:2000cv}. Comparisons with results obtained in several other theoretical
approaches can be found in Ref.~\cite{Geng:2009ys}. The convergence properties of the chiral series of the magnetic moments
for the $\Delta(1232)$
can be accessed looking separately at the LO and NLO contributions. Moreover we can separate the contribution coming from octet 
loops (second number) and the decuplet loops (third number):
\begin{eqnarray*}
&&\mu_{\Delta^{++}}=7.76(1-0.14-0.08)\hspace{0.15cm},\hspace{0.15cm}\mu_{\Delta^+}=3.88(1-0.18-0.09),\\
&&\mu_{\Delta^{0}}=0-0.31-0.05\hspace{0.15cm},\hspace{0.15cm} \mu_{\Delta^{-}}=-3.88(1-0.02-0.06).\label{Eq:convergenceD}
\end{eqnarray*} 
As in the case of the baryon octet magnetic moments the convergence properties are good. This is also true for the chiral
series of the magnetic moments of the rest of the decuplet members~\cite{Geng:2009ys}. Finally it is worth to mention that in
this approach not only the magnetic dipole (moment) of the spin-3/2 decuplet resonances has been investigated but also the
electric quadrupole and the magnetic octupole moments and the charge radius~\cite{Geng:2009ys}. 

In summary, we have studied the electromagnetic structure of the low-lying baryons in the approach of chiral perturbation
theory up to next-to-leading order. More precisely, we have succesfully described the magnetic moments of both octet and
decuplet baryons by using the covariant formulation within the EOMS renormalization prescription that is an extension of
$\overline{MS}$ and therefore fulfills the principles of analyticity as well as relativity. For the decuplet baryons we are able to provide
predictions for the magnetic dipole moments of the $\Delta(1232)$, in particular,  $\mu_{\Delta^{++}}=6.0(6)\;\mu_N$ and 
$\mu_{\Delta^{+}}=2.84(34)\;\mu_N$.

\section{Acknowledgments}

This work was partially supported by the  MEC grant  FIS2006-03438 and the European Community-Research Infrastructure
Integrating Activity Study of Strongly Interacting Matter (Hadron-Physics2, Grant Agreement 227431) under the Seventh Framework Programme of EU. L.S.G. acknowledges support from the MICINN in the Program 
``Juan de la Cierva''. J.M.C. acknowledges the same institution for a FPU grant.

\end{document}